\documentstyle[12pt,epsf]{iopart}

\begin{document}

\title[Fr\"{o}hlich-Coulomb model of high-temperature superconductivity]
{Fr\"{o}hlich-Coulomb model of high-temperature superconductivity 
and charge segregation in the cuprates.}

\author{A S Alexandrov\dag and P E Kornilovitch\ddag}

\address{\dag\ Department of Physics, Loughborough University, 
Loughborough LE11 3TU, UK}

\address{\ddag\ Hewlett Packard Labs, 1501 Page Mill Road 1L-1123, 
Palo Alto, CA 94304, USA}

\begin{abstract}

We introduce a generic Fr\"ohlich-Coulomb model of the oxides, which also
includes infinite on-site (Hubbard) repulsion, and describe a simple
analytical method of solving the multi-polaron problem in complex lattice
structures. Two particular lattices, a zig-zag ladder and a perovskite 
layer, are studied. We find that depending on the relative strength of the
Fr\"ohlich and Coulomb interactions these systems are either polaronic
Fermi (or Luttinger)-liquids, bipolaronic superconductors, or charge
segregated insulators. In the superconducting phase the carriers are
superlight mobile bipolarons. The model describes key features of the
cuprates such as their T$_c$ values, the isotope effects, the normal state
diamagnetism, pseudogap, and spectral functions measured in tunnelling and
photoemission. We argue that a low Fermi energy and strong coupling of
carriers with high-frequency phonons is the cause of high critical
temperatures in novel superconductors.

\end{abstract}

\submitto{\JPCM}

\pacs{71.27, 71.38.+i, 74.20.Mn}

\maketitle

There is overwhelming experimental \cite{guo,shen,tim,ega,mul3,mul2} and
theoretical \cite{alemot,dev,allen,gor} evidence for an exceptionally strong
electron-phonon (el-ph) interaction in the cuprates, which competes with 
electron correlations. In recent years, several publications addressed the
fundamental problem of competing el-ph and Coulomb interactions in the
framework of the so-called Holstein-Hubbard model 
\cite{bis2,feh,zey,bon2,aub2}, where both interactions are short-range 
(on-site).  The model describes well many properties of the insulating 
state of the cuprates, including antiferromagnetism, lattice distortions, 
and phase segregation. However, it could hardly account for the high value 
of the superconducting critical temperature \cite{ale2}. The mass of 
(bi)polaronic carriers in this model is very large in the relevant 
parameter region, and $T_{c}$ is suppressed below the kelvin scale.

In choosing the correct interaction for HTSC, we take into account that most
of the novel superconductors are doped insulators with highly polarizable
ionic lattices. The low density of mobile carriers is unable to screen
effectively the direct Coulomb electron-ion and electron-electron
interactions. The layered structure of the cuprates reduces screening even
further. Since the mobile carriers are confined to the copper-oxygen planes
their interaction with out-of-plane ions, such as the apical oxygens, is
particularly strong. A parameter-free estimate of the polaron binding
energy $E_{p}$ in the cuprates puts it at about 0.5 eV or larger 
\cite{alebra}. This strong long-range (Fr\"{o}hlich) el-ph interaction
necessarily leads to formation of small polarons. Such Fr\"{o}hlich small
polarons were first considered in reference \cite{yam}. Exact Monte-Carlo
simulations of the single polaron problem \cite{alekor} showed that a
long-range el-ph interaction effectively removes the difficulty with a large
polaron (and bipolaron) mass in the Holstein-type el-ph models. Indeed, the
polaron is heavy because it has to carry a lattice deformation with it which
is the same deformation that forms the polaron itself. Therefore, there 
exists a generic relation between $E_{p}$ and the renormalization of its 
mass: $m\propto \exp {(\gamma E_{p}/\omega )}$,
where $\omega $ is a characteristic phonon frequency and $\gamma \sim 1$ is
a numeric coefficient whose actual value depends on the radius of the
interaction. For a short-range el-ph interaction (Holstein) the {\em entire}
lattice deformation disappears and then forms in the new place when the
polaron moves between the nearest lattices sites. Therefore, $\gamma = 1$ 
and the polaron is very heavy for the characteristic cuprates values 
$E_{p}\sim 0.5$ eV and $\omega \sim 0.05$ eV. In the case of a long-range 
interaction, only a fraction of the total deformation changes every time 
the polaron moves and $\gamma $ could be as small as 0.25 \cite{ale2}. 
Clearly, this results in a dramatic mas reduction since 
$\gamma $ enters the exponent. Thus the effective mass could be 
$\leq 10\,m_{e}$ where a naive Holstein-like estimate would yield a 
huge mass $\sim 10,000\,m_{e}$. The above qualitative reasoning was 
fully confirmed by analytical \cite{ale2} and numerical 
(approximation-free Monte Carlo) \cite{alekor} studies of the double-chain 
and double-plane models with long-range el-ph interactions.  Later the 
single-polaron and bipolaron cases of the chain model were analyzed in 
more detail in references \cite{feh2} and \cite{bon}, respectively.  
These studies confirmed a much lower mass of both polaron and 
bipolaron in comparison to the Holstein-Hubbard limit.

Here we argue that a consistent theory of HTSC must include both the
long-range Coulomb repulsion between the carriers and the strong long-range
electron-phonon interaction. We propose an analytically solvable
multi-polaron model of high-temperature superconductivity that includes 
these realistic long-range interactions. From theoretical standpoint, the
long-range Coulomb repulsion is critical in ensuring that the carriers 
would not form large clusters. Indeed, in order to form stable {\em pairs} 
(bipolarons) the el-ph interaction has to be strong enough to overcome 
the Coulomb repulsion at short distances. Since the el-ph
interaction is long-range, there is a potential possibility for clustering.
We shall demonstrate that the inclusion of the Coulomb repulsion $V_{c}$
makes the clusters unstable. More precisely, there is a certain window of 
$V_{c}/E_{p}$ inside which the clusters are unstable but bipolarons
nonetheless form. In this parameter window the bipolarons are light and the
system is a superconductor with a high critical temperature. The bipolarons
repel each other and propagate in a band of about the same bandwidth as the
single-polaron bandwidth, in sharp contrast with all bipolaronic models
considered previously. At a weaker Coulomb interaction the system is a
charge segregated insulator. At a stronger Coulomb repulsion the system is a
polaron Fermi (or Luttinger) liquid. In the superconducting phase but close
to the clustering boundary, dynamical formation of short lived clusters or
stripes could be expected.

Our generic Fr\"{o}hlich-Coulomb model explicitly includes the electron
kinetic energy, the infinite-range Coulomb and electron-phonon interactions
as well as the lattice energy. The implicitly present infinite Hubbard $U$
prohibits double occupancy and removes the need to distinguish the fermionic
spin. Introducing spinless fermion operators $c_{{\bf n}}$ and phonon
operators $d_{{\bf m}\alpha }$, the model Hamiltonian is written as 
\begin{eqnarray}
\fl 
H = & - &\sum_{{\bf n\neq n^{\prime }}} T({\bf n-n^{\prime }})
c_{{\bf n}}^{\dagger } c_{{\bf n^{\prime }}} 
+ \sum_{{\bf n\neq n^{\prime }}}V_{c}({\bf n-n^{\prime }})
c_{{\bf n}}^{\dagger }c_{{\bf n}}c_{{\bf n^{\prime }}}^{\dagger }
c_{{\bf n^{\prime }}}  \nonumber \\
\fl
 & - &\omega \sum_{{\bf nm}}g_{\alpha }({\bf m-n})({\bf e}_{{\bf m}\alpha
}\cdot {\bf u}_{{\bf m-n}})c_{{\bf n}}^{\dagger }c_{{\bf n}}
(d_{{\bf m}\alpha }^{\dagger }+d_{{\bf m}\alpha }) +
\omega \sum_{{\bf m}\alpha }
\left(d_{{\bf m}\alpha }^{\dagger }d_{{\bf m}\alpha }
+\frac{1}{2}\right) .
\end{eqnarray}
We note that the el-ph term is written in real rather than momentum space.
This is more convenient in working with complex lattices. Here 
${\bf e}_{{\bf m}\alpha }$ is the polarization vector of $\alpha $th 
vibration coordinate at site 
${\bf m}$, ${\bf u}_{{\bf m-n}}\equiv ({\bf m-n})/|{\bf m-n}|$ is the unit 
vector in the direction from electron ${\bf n}$ to the ion ${\bf m}$, and 
$g_{\alpha }({\bf m-n)}$ is the dimensionless el-ph coupling function. 
[$g_{\alpha }({\bf m-n)}$ is proportional to the {\em force} acting 
between ${\bf m}$ and ${\bf n}$.] We assume that all the phonon modes 
are dispersionless with frequency $\omega $ and that the electrons do 
not interact with displacements of their own atoms, 
$g_{\alpha}(0)\equiv 0$. We also use $\hbar =1$ throughout the paper.

In general, the many-body model (1) is of considerable complexity.
However, we are interested in the limit of strong el-ph interaction. In
this case, the kinetic energy is a perturbation and the model can be grossly
simplified in a two-step procedure. On the first step, the Lang-Firsov
canonical transformation \cite{lan} is performed which diagonalizes the 
last three terms in equation (1).  Introducing
$S = \sum_{{\bf mn}\alpha } 
g_{\alpha }({\bf m-n})({\bf e}_{{\bf m}\alpha } \cdot 
{\bf u}_{{\bf m-n}})c_{{\bf n}}^{\dagger }
c_{{\bf n}}(d_{{\bf m}\alpha }^{\dagger} - d_{{\bf m}\alpha })$ 
one obtains the transformed Hamiltonian without an explicit el-ph term 
\begin{eqnarray}
\tilde{H} = e^{-S}He^{S} = 
& - & \sum_{{\bf n\neq n^{\prime }}}\hat{\sigma}_{{\bf %
nn^{\prime }}}c_{{\bf n}}^{\dagger }c_{{\bf n^{\prime }}}+\omega \sum_{{\bf m%
}\alpha }\left( d_{{\bf m}\alpha }^{\dagger }d_{{\bf m}\alpha }+\frac{1}{2}%
\right)   \nonumber \\
&+&\sum_{{\bf n\neq n^{\prime }}}v({\bf n-n^{\prime }})c_{{\bf n}}^{\dagger
}c_{{\bf n}}c_{{\bf n^{\prime }}}^{\dagger }c_{{\bf n^{\prime }}}-E_{p}\sum_{%
{\bf n}}c_{{\bf n}}^{\dagger }c_{{\bf n}}.
\end{eqnarray}
The last term describes the energy which polarons gain due to el-ph
interaction. $E_{p}$ is the familiar polaron (Franc-Condon) shift 
\begin{equation}
E_{p}=\omega \sum_{{\bf m}\alpha }g_{\alpha }^{2}({\bf m-n})({\bf e}_{{\bf m}%
\alpha }\cdot {\bf u}_{{\bf m-n}})^{2},
\end{equation}
which we assume to be independent of ${\bf n}$. $E_{p}$ is a natural measure
of the strength of the el-ph interaction. The third term in Eq.(2) is the
polaron-polaron interaction: 
\begin{equation}
v({\bf n-n^{\prime }})=V_{c}({\bf n-n^{\prime }})-
V_{{\rm pa}}({\bf n-n^{\prime }}),
\end{equation}
\begin{eqnarray}
V_{{\rm pa}}({\bf n-n^{\prime }}) &=&2\omega \sum_{{\bf m}\alpha }
g_{\alpha} ({\bf m-n})g_{\alpha }({\bf m-n^{\prime }})\times   \nonumber \\
&&({\bf e}_{{\bf m}\alpha }\cdot {\bf u}_{{\bf m-n}})({\bf e}_{{\bf m}\alpha
}\cdot {\bf u}_{{\bf m-n^{\prime }}}),
\end{eqnarray}
where $V_{{\rm pa}}$ is the inter-polaron {\em attraction} due to joint
interaction with the same vibrating atoms. Finally, the first term in
equation (2) contains the transformed hopping operator 
$\hat{\sigma}_{{\bf nn^{\prime }}}$: 
\begin{eqnarray}
\hat{\sigma}_{{\bf nn^{\prime }}} &=&T({\bf n-n^{\prime }})\exp \left[ \sum_{%
{\bf m}\alpha }\left[ g_{\alpha }({\bf m-n})({\bf e}_{{\bf m}\alpha }\cdot 
{\bf u}_{{\bf m-n}})\right. \right.   \nonumber \\
&-&\left. \left. g_{\alpha }({\bf m-n^{\prime }})({\bf e}_{{\bf m}\alpha
}\cdot {\bf u}_{{\bf m-n^{\prime }}})\right] (d_{{\bf m}\alpha }^{\dagger}
- d_{{\bf m}\alpha })\right] .  
\label{four}
\end{eqnarray}
At large $E_{p}/T({\bf n-n^{\prime }})$ this term is a perturbation. In the
first order of the strong coupling perturbation theory \cite{alemot}, $\hat{%
\sigma}_{{\bf nn^{\prime }}}$ should be averaged over phonons because there
is no coupling between polarons and phonons in the unperturbed Hamiltonian
[the last three terms in equation (2)]. For temperatures lower than 
$\omega $, the result is 
\begin{equation}
t({\bf n-n^{\prime }})\equiv \left\langle \hat{\sigma}_{{\bf nn^{\prime }}}
\right\rangle _{ph}=T({\bf n-n^{\prime }})\exp [-G^{2}({\bf n-n^{\prime }})],
\end{equation}
\begin{eqnarray}
G^{2}({\bf n-n^{\prime }}) &=&\sum_{{\bf m}\alpha }g_{\alpha }({\bf m-n})(%
{\bf e}_{{\bf m}\alpha }\cdot {\bf u}_{{\bf m-n}})\times   \nonumber \\
&&\left[ g_{\alpha }({\bf m-n})({\bf e}_{{\bf m}\alpha }\cdot {\bf u}_{{\bf %
m-n}})-g_{\alpha }({\bf m-n^{\prime }})({\bf e}_{{\bf m}\alpha }\cdot {\bf u}%
_{{\bf m-n^{\prime }}})\right] .  \label{seven}
\end{eqnarray}
By comparing equations (3) and (5) with equation (8), the mass 
renormalization exponents can be expressed via $E_{p}$ and $V_{\rm pa}$ 
as follows 
\begin{equation}
G^{2}({\bf n-n^{\prime }})=\frac{1}{\omega }
\left( E_{p}-\frac{1}{2}V_{{\rm pa}}
({\bf n-n^{\prime }})\right) .
\end{equation}
This results in a renormalized hopping term that represents the small
parameter of a strong coupling perturbation theory \cite{alemot}. The above
technical transformation is simple and has been described elsewhere 
\cite{alemot} together with a detailed description of the perturbation 
procedure.  The resulting model is purely polaronic in which phonons are 
``integrated out'' 
\begin{equation}
H_{p} = H_{0} + H_{\rm pert},
\end{equation}
\begin{equation}
H_{0} = -E_{p}\sum_{{\bf n}}c_{{\bf n}}^{\dagger }c_{{\bf n}}
+ \sum_{{\bf n\neq n^{\prime }}} v({\bf n-n^{\prime }})
c_{{\bf n}}^{\dagger }c_{{\bf n}}c_{{\bf n^{\prime }}}^{\dagger }
c_{{\bf n^{\prime }}},
\end{equation}
\begin{equation}
H_{{\rm pert}} = - \sum_{{\bf n\neq n^{\prime }}}
t({\bf n-n^{\prime }})c_{{\bf n}}^{\dagger }c_{{\bf n^{\prime }}}.
\end{equation}
When $V_{{\rm pa}}$ exceeds $V_{c}$ the full interaction becomes negative
and polarons form pairs. We emphasize that while the above formalism fails
in some regimes of the short-range el-ph models (for instance, the adiabatic
limit of the Holstein model), it is surprisingly accurate for long-range
el-ph interactions, as was demonstrated by comparing the analytical results
with exact quantum Monte Carlo data \cite{alekor}.  It makes theoretical 
analysis of even complex interactions and lattice geometries simple 
and instructive. But before proceeding to analyzing concrete lattices, 
let us elaborate on the physics behind the lattice sums in equations 
(3) and (5).

When a carrier (electron or hole) acts on an ion with a force ${\bf f}$, it
displaces the ion by some vector ${\bf x}={\bf f}/s$. Here $s$ is the ion's
force constant. The total energy of the carrier-ion pair is 
$-{\bf f}^{2}/(2s)$.  This is precisely the summand in equation (3) 
expressed via dimensionless coupling constants.  Now consider two carriers 
interacting with the {\em same} ion, see figure 1(a).  The ion displacement 
is ${\bf x}=({\bf f}_{1}+{\bf f}_{2})/s$ and the energy is 
$-{\bf f}_{1}^{2}/(2s)-{\bf f}_{2}^{2}/(2s)-({\bf f}_{1}\cdot {\bf f}_{2})/s$.
The last term here should be interpreted as an ion-mediated interaction 
between the two carriers. It depends on the scalar product of ${\bf f}_{1}$ 
and ${\bf f}_{2}$ and consequently on the relative positions of the carriers 
with respect to the ion. If the ion is an isotropic harmonic oscillator, 
as we assume in this paper, then the following simple rule applies. If the 
angle $\phi $ between ${\bf f}_{1}$ and ${\bf f}_{2}$ is less than $\pi /2$ 
then the polaron-polaron interaction is attractive, otherwise it is 
repulsive, see figure 1(b). The overall sign and 
magnitude of the interaction is given by the lattice sum in equation (5), 
evaluation of which is elementary.  Notice also that according to 
equation (9), an attractive interaction reduces the polaron mass (and 
consequently bipolaron mass), while repulsive interaction enhances the 
mass. Thus in our model the long-range character of the el-ph interaction 
serves the double purpose.  Firstly, it generates additional inter-polaron 
attraction because the distant ions have small angle $\phi $.  This 
additional attraction helps overcome the direct Coulomb repulsion between 
the polarons.  Secondly, the Fr\"{o}hlich interaction makes the bipolarons 
light leading to a high critical temperature.

The many-particle ground state of $H_{0}$ depends on the sign of the
polaron-polaron interaction, the carrier density, and the lattice geometry.
First we consider the zig-zag ladder, figure 2(a), assuming that all sites 
are isotropic two-dimensional harmonic oscillators. For simplicity, we 
also adopt the nearest-neighbour approximation for both interactions, 
$g_{\alpha}({\bf l}) \equiv g$, $V_{c}({\bf n}) \equiv V_{c}$, 
and for the hopping integrals, $T({\bf m}) = T_{NN} > 0$ for $l=n=m=a$, 
and zero otherwise.  Hereafter we set the lattice period $a=1$. 
There are four nearest neigbours in the ladder, $z=4$. 
The one-particle polaronic Hamiltonian takes the form 
\begin{equation}
\fl
H_{p}=-\sum_{n}\left(E_p [c^{\dagger}_{n}c_{n}+p^{\dagger}_{n}p_{n}] +
t^{\prime}[c^{\dagger}_{n+1}c_{n}+p^{\dagger}_{n+1}p_{n} + {\rm h.c.}] +
t[p^{\dagger}_{n} c_{n}+p^{\dagger}_{n-1}c_{n} + {\rm h.c.}] \right),
\label{eleven}
\end{equation}
where $c_{n}$ and $p_{n}$ are polaronic operators on the lower and upper
side of the ladder, respectively, see figure 2(b). 
Applying the general formulas (3), (5), and (9), we obtain 
$E_{p} = 4g^2\omega$, $t^{\prime}= T_{NN} \exp[-7E_{p}/(8\omega)]$, 
and $t = T_{NN} \exp[-3E_{p}/(4\omega)]$. 
Fourier transformation yields the one-particle spectrum 
\begin{equation}
E_{1}(k) = - E_{p} - 2t^{\prime}\cos(k) \pm t \cos(k/2).  
\label{twelve}
\end{equation}
Two overlapping polaronic bands have a combined width of $W = 4t^{\prime}+2t$.
The lower band has the bandwidth $W$ and the effective mass 
$m^{\ast}_{l} = 2/(4t^{\prime}+ t)$ near the bottom, while the upper band 
has the bandwidth $4t^{\prime}-2t$ and a heavier mass 
$m^{\ast}_{u} = 2/(4t^{\prime} - t)$.

Let us now place two polarons on the ladder. The nearest neighbour
interaction, equation (4), is found as $v = V_{c} - E_{p}/2$ if two 
polarons are on the different sides of the ladder, and 
$v = V_{c}-E_{p}/4$ if both polarons are on the same side. The attractive 
interaction is provided via the
displacement of the lattice sites which are the common nearest neighbours to
both polarons, under the condition that the angle $\phi $ between the
directions pointing from those sites to two polarons is less than $\pi/2$.
If $\phi \geq \pi/2$ the effective interaction of two polarons is repulsive.
There are two such nearest neighbours for the intersite bipolaron of the
type $A$ or $B$, figure 2(c), but there is only one common nearest neighbour 
for the bipolaron $C$, figure 2(d). When $V_{c} > E_{p}/2$, there are no 
bound states and the multi-polaron system is a 1D Luttinger liquid. 
However, when $V_{c} < E_{p}/2$ and consequently $v < 0$, the two polarons 
are bound into an intersite bipolaron of the type $A$ or $B$.

It is quite remarkable that the bipolaron tunnelling appears already in the
first order in polaron hopping $H_{\rm pert}$ as was anticipated in 
\cite{ale2}. This case is different from both the on-site bipolaron 
discussed a long time ago \cite{aleran}, and from the inter-site chain 
bipolaron discussed recently \cite{bon}, where the bipolaron tunnelling 
was of the second order in $t$.  Indeed, in the first order in 
$H_{\rm pert}$ one should consider only the lowest energy degenerate 
configurations $A$ and $B$ and discard the processes that involve all 
other configurations. The result of such a projection is a bipolaronic 
Hamiltonian 
\begin{equation}
\fl
H_{b} = \left( V_{c}- \frac{5}{2} E_{p} \right) \sum_{n} [A^{\dagger}_{n}
A_{n} + B^{\dagger}_{n} B_{n}] - t^{\prime}\sum_{n} [B^{\dagger}_{n} A_{n} +
B^{\dagger}_{n-1} A_{n} + {\rm h.c.} ],  \label{thirteen}
\end{equation}
where $A_{n} = c_{n}p_{n}$ and $B_{n} = p_{n}c_{n+1}$. 
Fourier transformation yields the bipolaron energy spectrum: 
\begin{equation}
E_{2}(k) = V_{c} - {\frac{5}{{2}}} E_p \pm 2t^{\prime} \cos (k/2).
\label{fourteen}
\end{equation}
There are two bipolaron bands with a combined width of $4t^{\prime}$. The
bipolaron binding energy is 
\begin{equation}
\Delta \equiv 2E_{1}(0) - E_{2}(0) = \frac{E_{p}}{2} - V_{c} - 2t -
4t^{\prime}.  \label{fifteen}
\end{equation}
The bipolaron mass near the bottom of the lowest band is $m^{\ast \ast} =
2/t^{\prime}$. Neglecting $t$ and $t^{\prime}$ relative to $E_p$ and $V_c$
we arrive at the following conclusion. When $V_c < E_p/2$, two polarons form
a bipolaron with effective mass $m^{\ast \ast} \approx (4+\exp{\frac{E_p}{%
8\omega}}) m^{\ast}_{l}$. The numerical coefficient $\frac{1}{8}$ ensures
that $m^{\ast \ast}$ remains of the order of $m^{\ast}$ even at large $E_p$.

In models with strong intersite attraction there is a possibility of
clasterization. In a way similar to the two-particle case above, the lowest
energy of $n$ polarons placed on the nearest neighbours of the ladder is
found as $E_{n} = (2n-3)V_c - \frac{6n-1}{4} E_{p}$ , for any $n\geq 3$.
There are {\em no} resonating states for $n$-polaron nearest neighbour
configuration if $n\geq 3$. Therefore there is no first-order kinetic energy
contribution to their energy. $E_n$ should be compared with the energy $%
E_{1} + (n-1)E_{2}/2$ of far separated $(n-1)/2$ bipolarons and a single
polaron for odd $n\geq 3$, or with the energy of far separated $n$
bipolarons for even $n\geq 4$. ``Odd'' clusters are stable at $V_{c} < \frac{%
n}{6n-10} E_{p}$, and ``even'' cluster are stable at $V_{c} < \frac{n-1}{%
6n-12} E_{p}$. Here we have neglected the kinetic energy of polarons and
bipolarons. As a result we find that bipolarons repel each other and single
polarons at $V_c > \frac{3}{8} E_p$. If $V_c$ is less than $\frac{3}{8} E_p$
then immobile bound clusters of three and more polarons could form. We would
like to stress that at distances much larger than the lattice constant the
polaron-polaron interaction is always repulsive \cite{ale2}, and the
formation of infinite clusters, stripes or strings is impossible 
\cite{alekab2}. Combining the condition of bipolaron formation and that 
of the instability of larger clusters we obtain a window of parameters 
\begin{equation}
\frac{3}{8} E_p < V_c < \frac{1}{2} E_p ,  
\label{nineteen}
\end{equation}
within which the ladder is a bipolaronic conductor. Outside this window the
ladder is either a charge segregated insulator (small $V_c$) or the
one-dimensional (1D) Luttinger liquid (large $V_c$).

Our consideration is directly related to doped cuprates. Here we consider a
two dimensional lattice of ideal octahedra that can be regarded as a
simplified model of the copper-oxygen perovskite layer, figure 3. The lattice
period is $a=1$ and the distance between the apical sites and the central
plane is $h=a/2=0.5$. All in-plane atoms, both copper and oxygen, are static
but apical oxygens are independent three-dimensional isotropic harmonic
oscillators. Because of poor screening the hole-apical interaction is
purely Coulombic, $g_{\alpha }({\bf m-n})=\kappa _{\alpha }/|{\bf m-n}|^{2}$, 
$\alpha =x,y,z$. To account for the experimental fact that $z$-polarized
phonons couple to the holes stronger than the others \cite{tim}, we choose 
$\kappa _{x}=\kappa _{y}=\kappa _{z}/\sqrt{2}$. The direct hole-hole
repulsion is $V_{c}({\bf n-n^{\prime }})=\frac{V_{c}/\sqrt{2}}
{|{\bf n-n^{\prime }}|}$ so that the repulsion between two holes in the NN
configuration is $V_{c}$. We also include the bare NN hopping $T_{NN}$, the
next nearest neighbor (NNN) hopping across copper $T_{NNN}$, and the NNN
hopping between the pyramids $T_{NNN}^{\prime }$. According to equation (3), 
the polaron shift is given by the lattice sum (after summation over
polarizations): 
\begin{equation}
E_{p}=2\kappa _{x}^{2}\omega \sum_{{\bf m}}
\left( \frac{1}{|{\bf m-n}|^{4}} + \frac{h^{2}}{|{\bf m-n}|^{6}}\right) 
= 31.15\kappa _{x}^{2}\omega ,
\label{televen}
\end{equation}
where the factor 2 accounts for the two layers of apical sites. [For
reference, Cartesian coordinates are ${\bf n}=(n_{x}+1/2,n_{y}+1/2,0)$, 
${\bf m}=(m_{x},m_{y},h)$; $n_{x},n_{y},m_{x},m_{y}$ being integers.] 
The polaron-polaron attraction is 
\begin{equation}
V_{{\rm pa}}({\bf n-n^{\prime }})=4\omega \kappa _{x}^{2}\sum_{{\bf m}}
\frac{h^{2}+({\bf m-n^{\prime }})\cdot 
({\bf m-n})}{|{\bf m-n^{\prime }}|^{3}|{\bf m-n}|^{3}}.  
\label{ttwelve}
\end{equation}
Performing the lattice summations for the NN, NNN, and NNN' configurations
one finds $V_{\rm pa} = 1.23\,E_{p}$, $0.80\,E_{p}$, and $0.82\,E_{p}$,
respectively. Substituting these results in equation (4) and (9) we obtain 
the full inter-polaron interaction: $v_{NN}=V_{c}-1.23\,E_{p}$, 
$v_{NNN} = \frac{V_{c}}{\sqrt{2}}-0.80\,E_{p}$, 
$v_{NNN}^{\prime } = \frac{V_{c}}{\sqrt{2}} - 0.82\,E_{p}$, and the mass 
renormalization exponents: $G_{NN}^{2} = 0.38(E_{p}/\omega )$, 
$G_{NNN}^{2} = 0.60(E_{p}/\omega )$ and 
$G_{NNN}^{\prime 2} = 0.59(E_{p}/\omega )$.

Let us now discuss different regimes of the model. At $V_c > 1.23 \, E_p$,
no bipolarons are formed and the systems is a polaronic Fermi liquid. The
polarons tunnel in the {\em square} lattice with NN hopping 
$t = T_{NN} \exp ( -0.38 E_p/\omega)$ and NNN hopping 
$t^{\prime}= T_{NNN} \exp ( -0.60 E_p/\omega)$. 
[Since $G^2_{NNN} \approx G^{\prime 2}_{NNN}$ one can neglect
the difference between NNN hoppings within and between the octahedra.] 
The single polaron spectrum is therefore 
\begin{equation}
E_1({\bf k})= - E_p - 2t^{\prime}[\cos k_x +\cos k_y] -
4t\cos(k_x/2)\cos(k_y/2) .  \label{tfifteen}
\end{equation}
The polaron mass is $m^{\ast} = 1/(t+2t^{\prime})$. Since in general 
$t > t^{\prime}$, the mass is mostly determined by the NN hopping amplitude 
$t$.  If $V_c < 1.23 \,E_p$ then intersite NN bipolarons form. 
The bipolarons tunnel in the plane via four resonating (degenerate) 
configurations $A$, $B$, $C$, and $D$, see figure 4. In the first order 
in $H_{\rm pert}$ one should retain only these lowest energy configurations 
and discard all the processes that involve configurations with higher 
energies. The result of such a projection is a bipolaron Hamiltonian 
\begin{eqnarray}
H_{b}&=&(V_{c}-3.23 \, E_{p}) \sum_{{\bf l}} 
[A^{\dagger}_{{\bf l}} A_{{\bf l}} + B^{\dagger}_{{\bf l}} B_{{\bf l}} 
+ C^{\dagger}_{{\bf l}} C_{{\bf l}} +
D^{\dagger}_{{\bf l}} D_{{\bf l}}] \cr & - & t^{\prime}\sum_{{\bf l}}
[A^{\dagger}_{{\bf l}} B_{{\bf l}} + B^{\dagger}_{{\bf l}} C_{{\bf l}} +
C^{\dagger}_{{\bf l}} D_{{\bf l}} + D^{\dagger}_{{\bf l}} A_{{\bf l}} + 
{\rm h.c.}]\cr 
& - &t^{\prime}\sum_{{\bf n}} [A^{\dagger}_{{\bf l-x}} B_{{\bf l}}
+ B^{\dagger}_{{\bf l+y}} C_{{\bf l}} + C^{\dagger}_{{\bf l+x}} D_{{\bf l}}
+ D^{\dagger}_{{\bf l-y}} A_{{\bf l}} + {\rm h.c.} ],  \label{tsixteen}
\end{eqnarray}
where ${\bf l}$ numbers octahedra rather than individual sites, 
${\bf x}=(1,0)$, and ${\bf y}=(0,1)$.  A Fourier transformation and 
diagonalization of a $4 \times 4$ matrix yields the bipolaron spectrum: 
\begin{equation}
E_{2}({\bf k}) = V_{c} - 3.23 E_p \pm 2t^{\prime}[\cos (k_x/2)\pm 
\cos(k_y/2)].  
\label{tseventeen}
\end{equation}
There are four bipolaronic subbands combined in the band of the width $%
8t^{\prime}$. The effective mass of the lowest band is $m^{\ast\ast} =
2/t^{\prime}$. The bipolaron binding energy is $\Delta = 1.23 E_p - V_c -
4(2t+t^{\prime})$. As in the ladder, the bipolaron moves already in the {\em
first} order in polaron hopping. This remarkable property is entirely due to
the strong on-site repulsion and long-range electron-phonon interaction that
leads to a non-trivial connectivity of the lattice. This situation is unlike
all other models studied previously. [Usually, the bipolaron moves only in
the second order in polaron hopping and therefore is very heavy.] In our
model, this fact combines with a weak renormalization of $t^{\prime}$
yielding a {\em superlight} bipolaron with mass $m^{\ast\ast} \propto
\exp(0.60\, E_p/\omega)$. We recall that in the Holstein model $m^{\ast\ast}
\propto \exp(2 E_p/\omega)$. Thus the mass of the Fr\"ohlich bipolaron
scales approximately as a {\em cubic root} of that of the Holstein one.

At even stronger el-ph interaction, $V_c < 1.16 E_p$, NNN bipolarons become
stable. More importantly, holes can now form 3- and 4-particle clusters. The
dominance of the potential energy over kinetic in Hamiltonian (10)
enables us to readily investigate these many-polaron cases. Three holes
placed within one oxygen square have four degenerate states with energy 
$2(V_c-1.23E_p)+\frac{V_c}{\sqrt{2}}-0.80E_p$. The first-order polaron
hopping processes mix the states resulting in a ground state linear
combination with energy $E_3 = 2.71V_c -3.26 E_p - \sqrt{4t^2+t^{\prime 2}}$. 
It is essential that between the squares such triads could move only in
higher orders in polaron hopping. In the first order, they are immobile. A
cluster of four holes has only one state within a square of oxygen atoms.
Its energy is $E_4 = 4(V_c-1.23E_p)+2(\frac{V_c}{\sqrt{2}}-0.80E_p) = 5.41
V_c - 6.52 E_p$. This cluster, as well as all the bigger ones, are also
immobile in the first order of polaron hopping. We conclude that at 
$V_c < 1.16 \, E_p$ the system quickly becomes a charge segregated insulator.

The fact that within the window $1.16E_{p}<V_{c}<1.23E_{p}$ there are no
three or higher polaron bound states, means that bipolarons repel each
other. The system is effectively the charged Bose-gas which is a well known
superconductor \cite{alemot}. The superconductivity window that we have
found, is quite narrow (see figure 5). This indicates that the 
superconducting state in such systems is a subtle phenomenon which requires 
a fine balance between electronic and ionic interactions. Too strong el-ph 
interaction leads to clustering, while too weak interaction cannot bind the 
carriers and the superconductivity is at best of BCS type.  These 
considerations may provide additional insight into the uniqueness of one 
particular structure, the copper-oxygen perovskite layer, for HTSC. 
It also follows from our model that superconductivity should be very 
sensitive to any external factor that affects the balance between $V_{c}$ 
and $E_{p}$. For instance, pressure changes the octahedra geometry and 
hence $E_{p}$ and $V_{\rm pa}$.  Chemical doping enhances internal 
screening and consequently reduces $E_{p}$.

We now assume that the superconductivity condition is satisfied and show
that our Fr\"ohlich-Coulomb model possesses many key properties of the
underdoped cuprates. The bipolaron binding energy $\Delta$ should manifest
itself as a normal state pseudogap with size of approximately half of 
$\Delta $ \cite{alemot}. Such a pseudogap was indeed observed in many
cuprates. In contrast with the BCS superconductor, the symmetry of the
pseudogap might differ from the symmetry of the superconducting order
parameter, which depends on the bipolaronic band dispersion. The symmetry 
of the order parameter was
found to be $d$-wave \cite{ale4}, while the former is an anisotropic
s-wave, in accordance with many experimental observations. There should be a
strong isotope effect on the (bi)polaron mass because $t,t^{\prime}\propto
\exp(-{\rm const} \sqrt{M}) $. Therefore the replacement of O$^{16}$ by 
O$^{18}$ increases the carrier mass \cite{ale3}. Such an effect was observed
in the London penetration depth of the isotope-substituted samples \cite{guo}.
The mass isotope exponent, $\alpha_m=d\ln m^{\ast\ast}/d\ln M$, was found
to be as large as $\alpha_m = 0.8$ in La$_{1.895}$Sr$_{0.105}$CuO$_4$. Our
theoretical exponent is $\alpha_m=0.3 E_{p}/\omega$, so that the bipolaron
mass enhancement factor is $\exp(0.6E_p/\omega) \simeq 5$ in this material.
With the bare hopping integral $T_{NNN}=0.2$ eV we obtain the in-plane
bipolaron mass $m^{\ast \ast} \simeq 10 \, m_e$. Calculated with this value 
the in-plane London penetration depth, $\lambda_{ab}=[m^{\ast \ast}/8\pi
ne^{2}]^{1/2}\simeq 316$ nm ($n$ the hole density) agrees well with the
measured one $\lambda_{ab} \simeq 320$ nm. Taking into account the c-axis
tunnelling of bipolarons, the critical temperature of their Bose-Einstein
condensation can be expressed in terms of the experimentally measured
in-plane and c-axis penetration depths, and the in-plane Hall constant 
$R_{H} $ as $T_c \approx 1.64 (eR_H/\lambda_{ab}^4\lambda_{c}^2)^{1/3}$. 
Here $T_c$, $eR_{H}$, and $\lambda $ are measured in K, cm$^3$ and cm,
respectively \cite{alekab}. Using the experimental $\lambda_{ab} = 320$ nm, 
$\lambda_{c} = 4160$ nm, and $R_H = 4 \times 10^{-3}$ cm$^3$/C (just above 
$T_c$), one obtains $T_c=31$ K in striking agreement with the experimental
value $T_c=30$ K. The recent observation of the normal state diamagnetism in
La$_{2-x}$Sr$_{x}$CuO$_{4}$ \cite{nat} also fits well the prediction of the
bipolaron theory \cite{alekabden}. Many other features of the bipolaronic
(super)conductor, e.g., the unusual upper critical field, electronic
specific heat, optical, ARPES and tunnelling spectra match those in the
cuprates (for a recent review, see reference \cite{aleedw}).

Finally, we show that the Fermi energy in all novel superconductors is
surprisingly low, of the order or even smaller than the most essential
optical phonon energy. The band structure of the cuprates is
quasi-two-dimensional with a few degenerate hole pockets. Applying the
parabolic approximation for the band dispersion one obtains the renormalized
(polaronic) Fermi energy as 
\begin{equation}
E_{F}={\frac{\pi n_{i}d}{{m_{i}^{\ast }}}},  
\label{twentyone}
\end{equation}
where $d$ is the interplane distance, and $n_{i},m_{i}^{\ast }$ are the
density of holes and their effective mass in each of the hole subbands $i$
renormalized by the electron-phonon (and electron-electron) interactions.
One can express the renormalized band-structure parameters through the
in-plane London penetration depth at $T=0$, measured experimentally: 
\begin{equation}
\frac{1}{\lambda _{H}^{2}}=4\pi e^{2}\sum_{i}\frac{n_{i}}{m_{i}^{\ast }}.
\label{twentytwo}
\end{equation}
As a result, one obtains the {\em parameter-free} expression for the Fermi
energy as 
\begin{equation}
E_{F} = {\frac{d}{{4ge^{2}\lambda _{H}^{2}}}},  
\label{twentythree}
\end{equation}
where $g$ is the degeneracy of the spectrum. The degeneracy $g$ in the
cuprates may depend on doping. In underdoped cuprates one expects 4 hole
pockets inside the Brillouin zone (BZ) due to the Mott-Hubbard gap. If the
hole band minima are shifted with doping to BZ boundaries, the spectrum will
be two-fold degenerate, so that $g\geq 2$ in cuprates. Because equation 
(\ref{twentythree}) does not contain any other band-structure parameters, 
the estimate of $E_{F}$ using this equation does not depend very much on the
parabolic approximation for the band dispersion. Generally, the ratios $n/m$
in equations (\ref{twentyone}) and (\ref{twentytwo}) are not necessary the 
same.  The ``superfluid'' density in equation (\ref{twentytwo}) might be 
different from the total density of delocalized carriers in equation 
(\ref{twentyone}). However, in a translational invariant system they must 
be the same \cite{leg}. This is true even in the extreme case of a pure 
two-dimensional superfluid, where quantum fluctuations might be important. 
One can obtain a reduced value of the zero temperature superfluid density 
only in the dirty limit $l\ll \xi (0) $ where $\xi (0)$ is the 
zero-temperature coherence length. The latter was measured directly in 
cuprates as the size of the vortex core. It is about 10 \AA\ or even less. 
On the contrary, the mean free path was found surprisingly large at low 
temperatures, $l\sim $ 100-1000 \AA. Hence, the cuprates are in the clean 
limit, $l\gg \xi (0)$, so that the parameter-free expression for $E_{F}$, 
equation (\ref{twentythree}), is perfectly applicable.

Equation (26) yields E$_F \leq 100$ meV for the cuprates, especially if the
degeneracy $g\geq 2$ is taken into account. A few examples are 
La$_{1.85}$Sr$_{0.15}$CuO$_{4}$ (T$_c=$ 37 K, $\lambda_H = 240$ nm 
\cite{alekab}, $d = 0.66$ nm) with $gE_F = 77$ meV, 
YBa$_{2}$Cu$_{3}$O$_{6.92}$ (T$_c = 91.5$ K , $\lambda_H = 186$ nm 
\cite{alekab}, $d = 0.43$ nm) with $gE_F = 84$ meV. That should be compared 
with the characteristic phonon frequency, which is estimated as the plasma 
frequency of oxygen ions, $\omega =(4\pi Z^{2}e^{2}N/M)^{1/2} $. One 
obtains $\omega = 84$ meV with $Z=2$, $N=6/V_{cell}$, $M=16$ a.u. for 
YBa$_{2}$Cu$_{3}$O$_{6}$. Here $V_{cell}$ is the volume of the chemical 
unit cell. The low Fermi energy $E_{F}\leq \omega $ is a serious problem 
for the Migdal-Eliashberg approach. The non-crossing diagrams cannot be 
treated as vertex {\em corrections} because $\omega /E_{F}\geq 1$, since 
they are comparable to the standard terms. On the contrary, the estimate 
of $E_F $ supports further the nonadiabatic (bi)polarons as the 
(super)carriers in high-T$_c$ superconductors.

In conclusion, we have introduced a realistic multi-polaron model of
high-temperature superconductivity with the strong Fr\"{o}hlich and Coulomb
long-range interactions. We have described a simple procedure of calculating
polaron and bipolaron masses, identified and quantitatively analyzed a new
resonance mechanism of bipolaron mass reduction, and found the conditions
for clustering of holes and the window for their high-T$_{c}$
superconductivity.  The model possesses a rich phase diagram in the
coordinates of the inter-site Coulomb repulsion $V_{c}$ and the polaronic
(Franck-Condon) level shift $E_{p}$, see figure 5. The ground state is a 
polaronic Fermi (or Luttinger) liquid, for the strong Coulomb repulsion, 
bipolaronic high-temperature superconductor for the intermediate Coulomb 
repulsion, and the charge-segregated insulator for the weak repulsion. 
Remarkably, the inter-site bipolarons in the superconducting phase are 
``superlight'', propagating coherently with about the same mass as single 
polarons. In our model the bipolarons tunnel already in the first order 
in polaron tunnelling which results in the bipolaron mass scaling 
linearly with the polaron hopping integral. Many properties of the model 
in the superconducting phase match those of the cuprates. We argue that 
a surprisingly low Fermi energy and the strong unscreened coupling 
of carriers with high frequency optical phonons is the origin of high 
temperature superconductivity.

This work has been supported by EPSRC UK (grant R46977), by the Leverhulme
Trust (grant F/00261/H), and by DARPA.

\begin{figure}[t]
\begin{center}
\leavevmode
\hbox{
\epsfxsize=14.cm
\epsffile{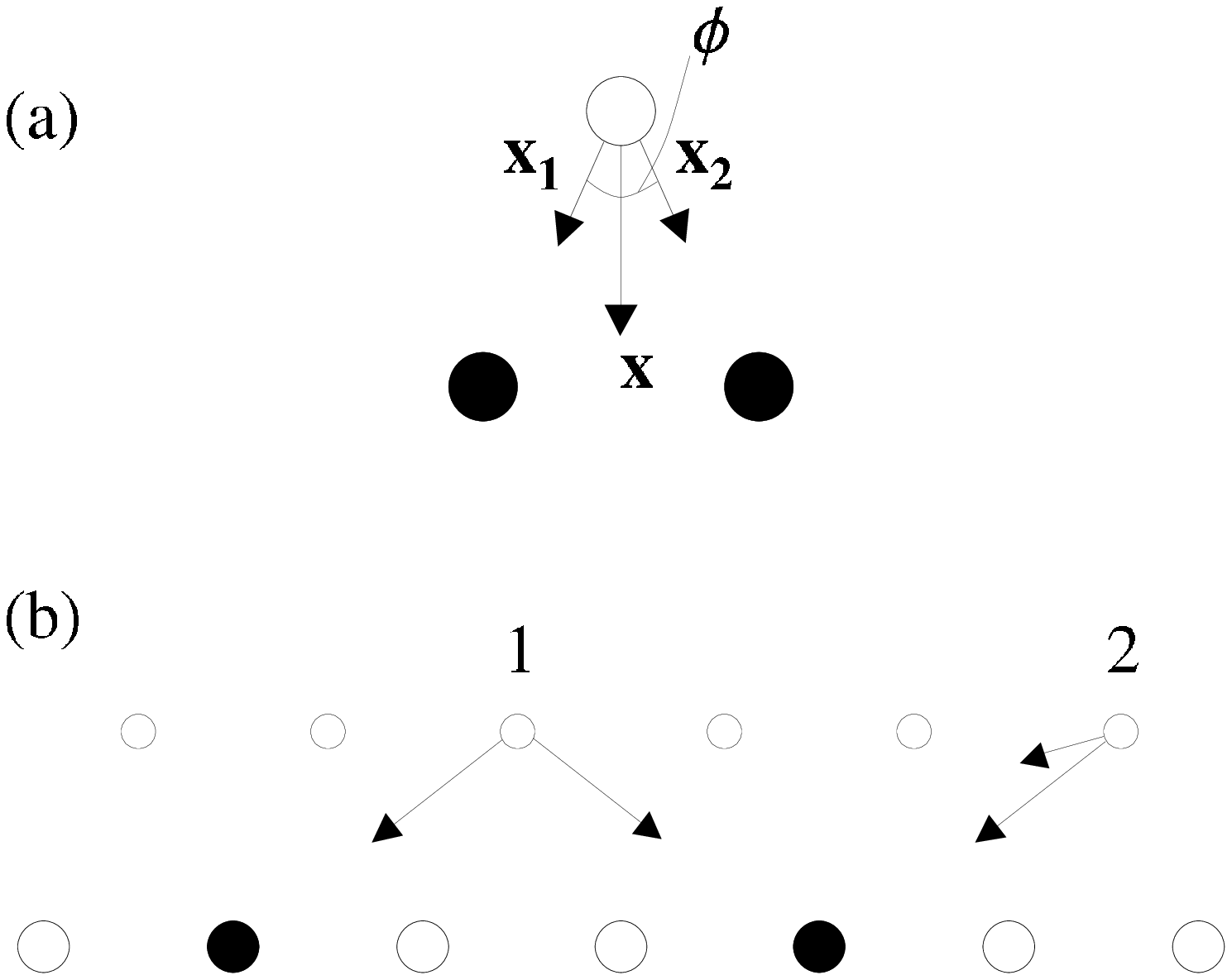}
}
\end{center}
\caption{
The mechanism of the polaron-polaron interaction. (a) Together,
the two polarons (solid circles) deform the lattice more effectively than
separately. An effective attraction occurs when the angle $\phi$ between 
${\bf x}_1$ and ${\bf x}_2$ is less than $\pi/2$. 
(b) A mixed situation. Atom {\bf 1} results in repulsion between two 
polarons while atom {\bf 2} results in attraction.
}
\label{fig1}
\end{figure}

\newpage

\begin{figure}[t]
\begin{center}
\leavevmode
\hbox{
\epsfxsize=14.cm
\epsffile{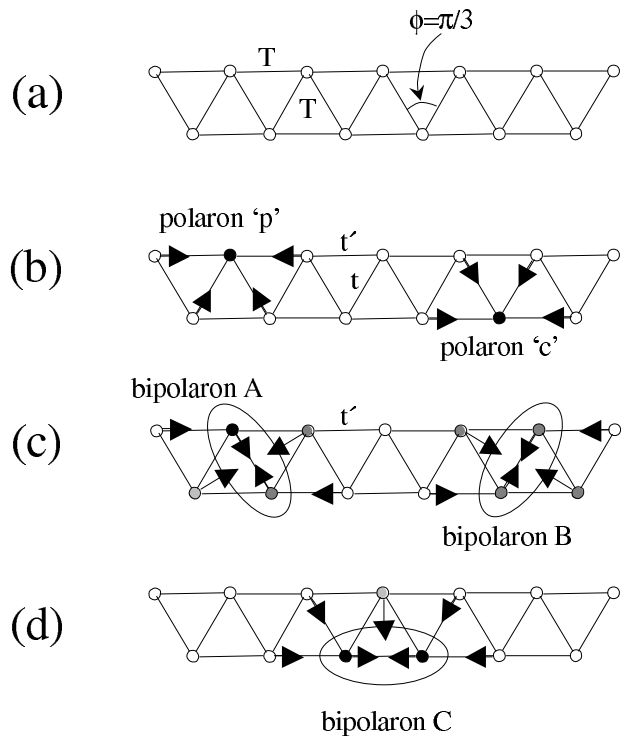}
}
\end{center}
\caption{
One-dimensional zig-zag ladder. (a) Initial ladder with the bare
hopping amplitude $T$. (b) Two types of polarons with their respective
deformations. (c) The two degenerate bipolaron configurations $A$ and $B$. 
(d) A different bipolaron configuration C which energy is higher than 
that of $A$ and $B$.
}
\label{fig2}
\end{figure}

\newpage

\begin{figure}[t]
\begin{center}
\leavevmode
\hbox{
\epsfxsize=14.cm
\epsffile{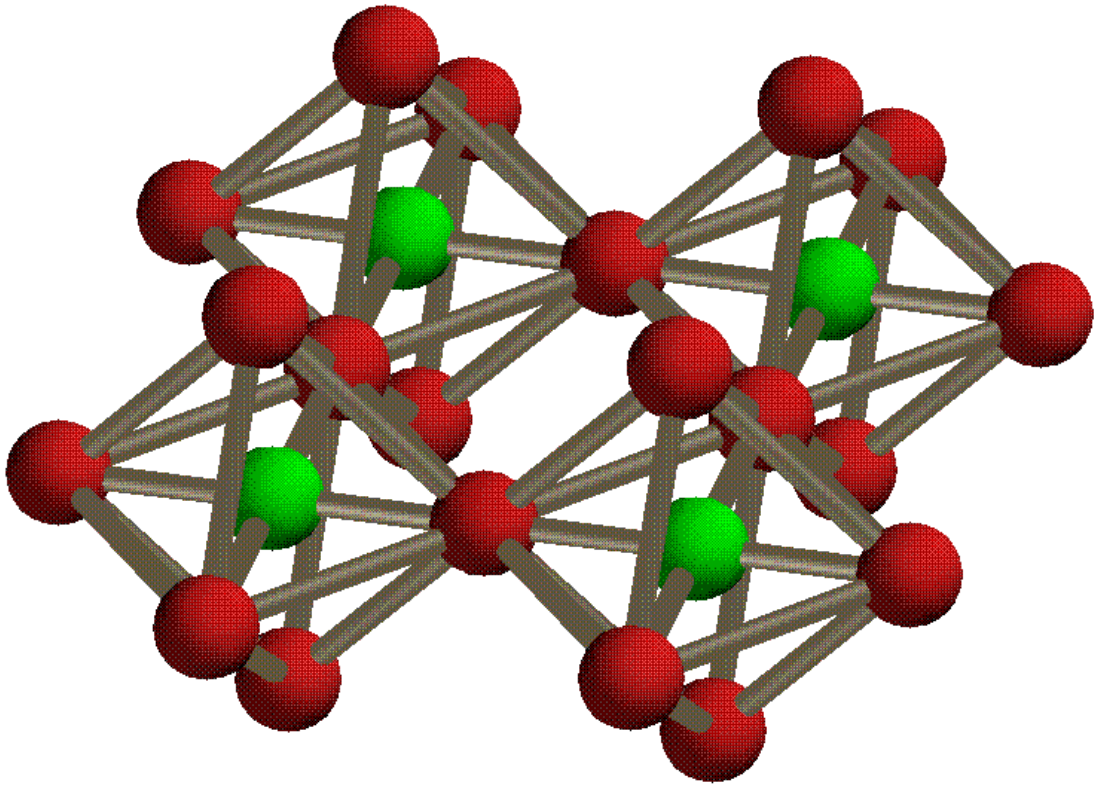}
}
\end{center}
\caption{
A fragment of the perovskite layer.
}
\label{fig3}
\end{figure}

\newpage

\begin{figure}[t]
\begin{center}
\leavevmode
\hbox{
\epsfxsize=14.cm
\epsffile{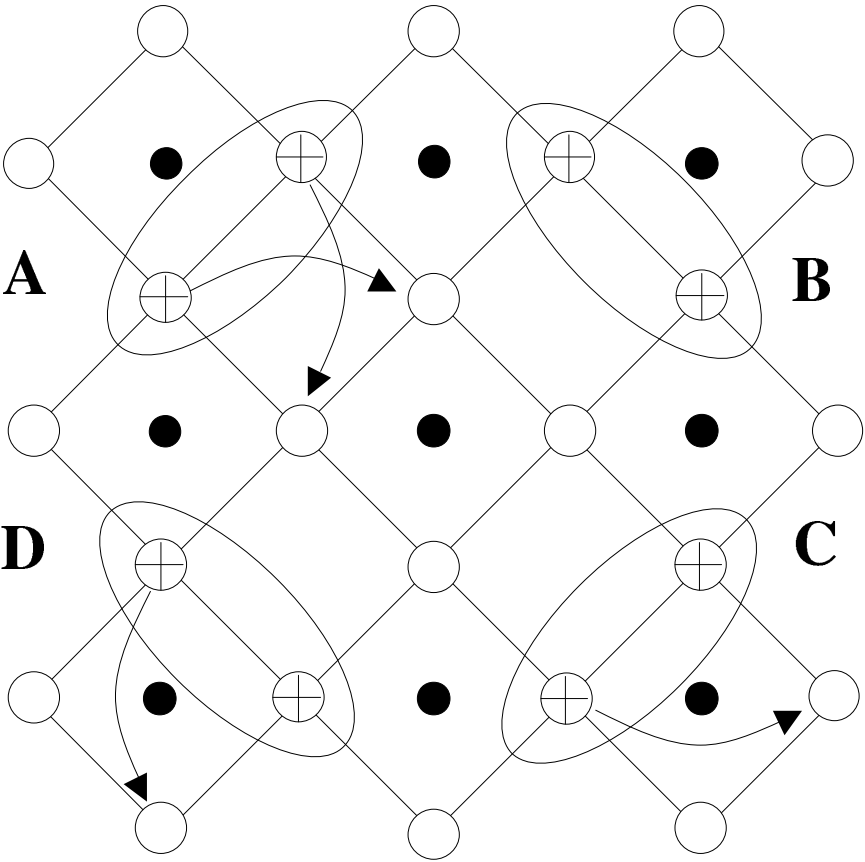}
}
\end{center}
\caption{
Four degenerate bipolaron configurations $A$, $B$, $C$, and $D$. 
Some single-polaron hoppings are indicated by arrows.
}
\label{fig4}
\end{figure}

\newpage

\begin{figure}[t]
\begin{center}
\leavevmode
\hbox{
\epsfxsize=14.cm
\epsffile{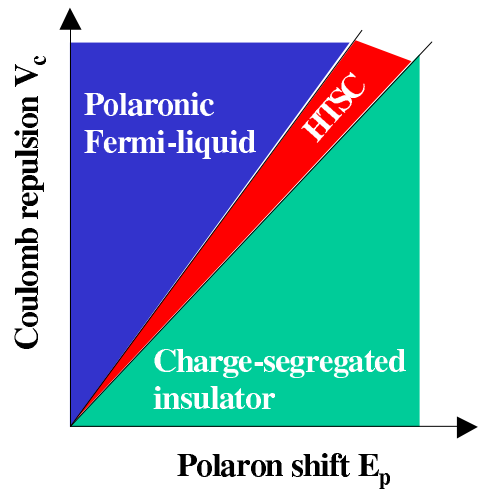}
}
\end{center}
\caption{
Phase diagram of the Fr\"{o}hlich-Coulomb model. The model is a
polaronic Fermi liquid for the strong Coulomb repulsion, bipolaronic
high-temperature superconductor (HTSC) for the intermediate Coulomb
repulsion, and the charge-segregated insulator for the weak repulsion.
}
\label{fig5}
\end{figure}

\end{document}